\documentclass[twocolumn,showpacs,superscriptaddress,preprintnumbers,amsmath,amssymb,prc]{revtex4}

\usepackage{graphicx}
\usepackage{dcolumn}
\usepackage{bm}
\usepackage{indentfirst} 
\usepackage{color}
\usepackage{CJK}
\usepackage{booktabs}
\usepackage[colorlinks,linkcolor=blue,urlcolor=blue,citecolor=blue]{hyperref}

\begin{document}
\begin{CJK}{UTF8}{gbsn}

\title{Effects of neutron-skin thickness on direct hard photon emission from reactions induced by the neutron-rich projectile $^{50}$Ca}

\author{S. S. Wang(王闪闪)}
\affiliation{Key Laboratory of Nuclear Physics and Ion-beam Application (MOE), Institute of Modern Physics, Fudan University, Shanghai 200433, China}

\author{Y. G. Ma(马余刚)}\thanks{Corresponding author: mayugang@fudan.edu.cn}
\affiliation{Key Laboratory of Nuclear Physics and Ion-beam Application (MOE), Institute of Modern Physics, Fudan University, Shanghai 200433, China}
\affiliation{Shanghai Research Center for Theoretical Nuclear Physics, NSFC and Fudan University, Shanghai 200438, China}

\author{D. Q. Fang(方德清)}
\affiliation{Key Laboratory of Nuclear Physics and Ion-beam Application (MOE), Institute of Modern Physics, Fudan University, Shanghai 200433, China}
\affiliation{Shanghai Research Center for Theoretical Nuclear Physics, NSFC and Fudan University, Shanghai 200438, China}

\author{X. G. Cao(曹喜光)}
\affiliation{Shanghai Advanced Research Institute, Chinese Academy of Sciences, Shanghai 201210, China}

\date{\today}

\begin{abstract}
Direct hard photon emissions from incoherent proton-neutron bremsstrahlung in collisions of the neutron-rich projectile $^{50}$Ca with $^{12}$C and $^{40}$Ca targets are simulated in the framework of the isospin-dependent quantum molecular dynamics (IQMD) model. By adjusting the diffuseness parameter of neutron density in the droplet model to obtain different neutron skin thicknesses for $^{50}$Ca, the effects of neutron skin thickness on direct hard photon emission are investigated via several probes. The results show that more direct hard photons are produced with increasing neutron skin thickness in peripheral collisions. Meanwhile, we find that the multiplicity yield ratio $R_{cp}(\sigma_{\gamma})$ between central collisions and peripheral collisions as well as the rapidity dependence of multiplicity for direct hard photons are sensitive to neutron skin thickness. The results indicate that direct hard photon emission can be taken  as an experimental observable to extract information on neutron skin thickness.
\end{abstract}

\maketitle

\section{Introduction}
\label{introduction}
The neutron skin of a nucleus，an important fundamental property，has attracted much attention in low-intermediate energy heavy-ion physics and nuclear astrophysics \cite{Shen,Tsang,Steiner,Li,PRL_new}. Very recently the significance of the neutron skin was also recognized in the relativistic heavy-ion collision community \cite{LiHL,Ham,Pau,De}. The neutron skin is usually defined as the difference between the root-mean-squared (rms) radii of neutrons and protons, i.e.,  $\delta_{np} = \langle r^{2}_n\rangle^{1/2} -\langle r^{2}_p\rangle^{1/2}$, and its formation in a nucleus depends on the balance between the inward pressure of the surface tension on excess neutrons on the edge of the nucleus and outward degeneracy pressure from excess neutrons within the core of the nucleus. Physically, it is closely related to the nuclear equation of state (EOS), especially for the symmetry energy [$E_{sym}(\rho)$] term \cite{Li,Chen2005PRC,Li2,XuJ}. A large amount of theoretical studies based on mean-field theories \cite{Chen2005PRC,Chen2010PRC,Yoshida2004PRC} and droplet-type models \cite{Centelles2009PRL,Danielewicz2003NPA} have pointed out that the neutron skin thicknesses of neutron-rich nuclei correlates linearly with slope parameter of $E_{sym}(\rho)$ at saturation density. Great efforts using different experimental probes, including proton elastic scattering \cite{Starodubsky1994PRC}, x-ray emission from antiprotonic atoms \cite{Trzcinska2001PRL}, parity-violating electron scattering \cite{Tarbert2014PRL}, isovector spin-dipole resonances \cite{Krasznahorkay1999PRL} and pygmy dipole resonances \cite{Klimkiewicz2007PRC}, to measure the neutron skin of a neutron-rich nucleus have been done. Therefore，to extract the information of neutron skin thickness with higher accuracy is of crucial importance for enriching our knowledge of neutron-rich matter and exploring the EOS to a higher nucleon  density, which helps us to understand many important properties of compressed nuclear matter and even of neutron stars.

Experimentally, proton rms radius can be probed to a very high accuracy with electromagnetic interaction ~\cite{Angeli2013nskinADNDT}. In contrast, it is considerably difficult to perform measurements of the neutron (weak charge) density distribution with enough precision and to make detailed  comparison with that of protons \cite{Gaidarov2020nskinNPA,Novario2021nskinaxXiv}. Recent high-precision measurements of neutron skin thickness for $^{208}$Pb by the PREX experiment \cite{Abrahamyan2012nskinPRL,Adhikari2021nskinPRL} and $^{48}$Ca by the CREX experiment\cite{Horowitz2014nskinEPJ} make it possible to carry out a precise measurement of the neutron radius. But one-part-per-million parity-violating asymmetry hinders the precise measurement of the neutron radius for short-lived isotopes \cite{Donnelly1989nskinNPA}. Therefore, more indirect experimental observables that are sensitive to neutron skin thickness are still very welcome. 

Using the isospin-dependent quantum molecular dynamics (IQMD) model with the different neutron and proton density distributions in the phase-space initialization, Refs.~ \cite{Sun2011nskin, Sun2011nskin2} proposed that the yield ratios of neutron to proton $[R(n/p)]$ can be taken  as an experimental observable to extract the neutron skin thickness. Then Ref.~\cite{Dai2014nskinPRC} indicated that the yield ratios of $^{3}$H to $^{3}$He [R(t/$^{3}$He)] could be treated as another possible experimental observable to extract the proton skin thickness. Recently, Refs.~\cite{Yan2019skinNST,Yan2019skinPRC} have also supported that the two above probes are sensitive to neutron skin thickness. Additionally, it was proposed to extract the proton rms radii $R_p$ \cite{Ozawa} and then deduce the neutron skin from charge-changing cross sections \cite{LiXF}. Moreover, Ref.~ \cite{Dai2015nskinPRC} pointed out that both the isoscaling parameter $\alpha$ and the mean value of $N/Z$ [$N (Z)$ is neutron (proton) number] of projectile-like fragments (PLFs) have a linear dependence on neutron skin thickness. However, compared to nucleons, light fragments, and PLFs produced in the reaction, hard photons have a considerable advantage since they are not being disturbed by the final-state interactions. Therefore, hard photons provide a clean probe of the reaction dynamics and  deliver an unperturbed picture of the emitting source \cite{Schutza1998photonNPA,Liu2008photonPLB, Ma2012photonPRC,Deng2018magneticEPJA,Yong2017photonPRC,Nifenecker1990photonARPNS,Shi1,Shi2}.

Pioneering experimental works \cite{Marques1994PRL,Badala1995PRL} extended the intensity interferometry method to hard photons in heavy-ion collisions at intermediate energies and it was clearly found in a model independent way that the hard photons come from a very small and transient sources \cite{Badala1995PRL}, whose space-time characteristic is consistent with the incoherent first-channel nucleon-nucleon collisions in the projectile-target overlap zone. So far, many experimental \cite{Tam1988photonPRC,Kwato1988photonNPA,Stevenson1986PRL,Grosse1986photonEL,Bertholet1987photonNPA,Migneco1993photonPLB} and theoretical \cite{Remington1986photonPRL,Ko1985photonPRC,Ko1986photonPRC,Nakayama1986photonPRC,Niita1988photonNPA,Bauer1986photonPRC,Cassing1986photonPLB,Khoa1991photonNPA} works have been done to understand the hard photon production mechanism in heavy-ion collisions. Nice reviews of hard-photon production are  given by Refs.~\cite{Cassing,Bonasera}, where  energetic particles as probes of the first stage of the reaction are deeply discussed. Based on these studies, it has been pointed out that hard photons are emitted from two distinct sources, i.e., direct hard photon and thermal hard photon sources, in space and time according to experimental evidence and Boltzmann-Uehling-Ulenbeck model calculations ~\cite{Mart1995photonPLB}. Direct hard photons stem from the first compression phase in the early stage of the reaction, which accounts for the dominant contribution. Thermal photons are produced from a thermalized source during the later stage of the reaction.  

In the present work, the IQMD model takes into account the in-medium effects by introducing the in-medium nucleon-nucleon cross section in the process of two-body collisions. A channel of incoherent proton-neutron bremsstrahlung collisions is embedded into the model. In recent calculations, we performed a comparison with experimental data and confirmed reliability of the method and model~\cite{Wang2020photonPRC}. Moreover, considering that direct hard photons originating from the earlier stage of the reaction may retain some evidence of the initial projectile,  here we shall focus on the effects of neutron skin thickness on direct hard photons emission from a reaction induced by the neutron-rich projectile $^{50}$Ca.

The paper is arranged as follows: In Sec. \ref{modelformalism}, a brief review of the IQMD model and the formula of hard photon production probability are given. Results and discussion are described in Sec.~\ref{resultsanddiscussion}, where the sensitivities of several probes of neutron skin thickness are checked and discussed via direct hard photons, including yield and yield ratio, and rapidity dependence of the multiplicity of direct hard photons. Finally, Sec.~\ref{summary} gives a summary.

\section{MODEL AND FORMALISM}
\label{modelformalism}
\subsection{Brief review of the IQMD model}

The isospin-dependent quantum molecular dynamics  model is a many-body theory which was developed from the standard QMD model by introducing isospin degrees of freedom into three components of the dynamics in heavy-ion collisions at intermediate energy, namely, the mean field, two-body collisions, and Pauli blocking ~\cite{Aichelin1991qmdPR,Hartnack1998qmdEPJA,Chen1988iqmdPRC,Zhang2018iqmdNST,Wang2020photonEPJA,Wang2019corrPRC,Yan2019skinNST,Yan2019skinPRC,Yan2019isospinNST,Feng2018qmdNST,Sun2011nskin,Dai2014nskinPRC,Dai2015nskinPRC}. In the model, each nucleon state is represented by a Gaussian wave function with width $L$ = 2.16 $fm^{2}$,
\begin{equation}\label{wavesingle}
\phi_{i}(\textbf{r},t) = \frac{1}{(2\pi L)^{3/4}} \exp[ -\frac{(\textbf{r}-\textbf{R}_{i})^{2}}{4 L}+\frac{i\textbf{P}_{i}\cdot \textbf{r}}{\hbar}],
\end{equation}
where $ \textbf{R}_{i} $ and $ \textbf{P}_{i} $ are the centers of position and momentum of the $i$-th wave packet, respectively. For a $N$-nucleon system, the total wave function $\Phi(\textbf{r},t)$ that evolves with time $t$ is given by a direct product of these nucleons' wave functions,
\begin{equation}\label{wavetotal}
\Phi(\textbf{r},t) = \prod^{N}_{i}\phi_{i}(\textbf{r},t).
\end{equation}

\begin{figure}
    \includegraphics[width=8.6cm]{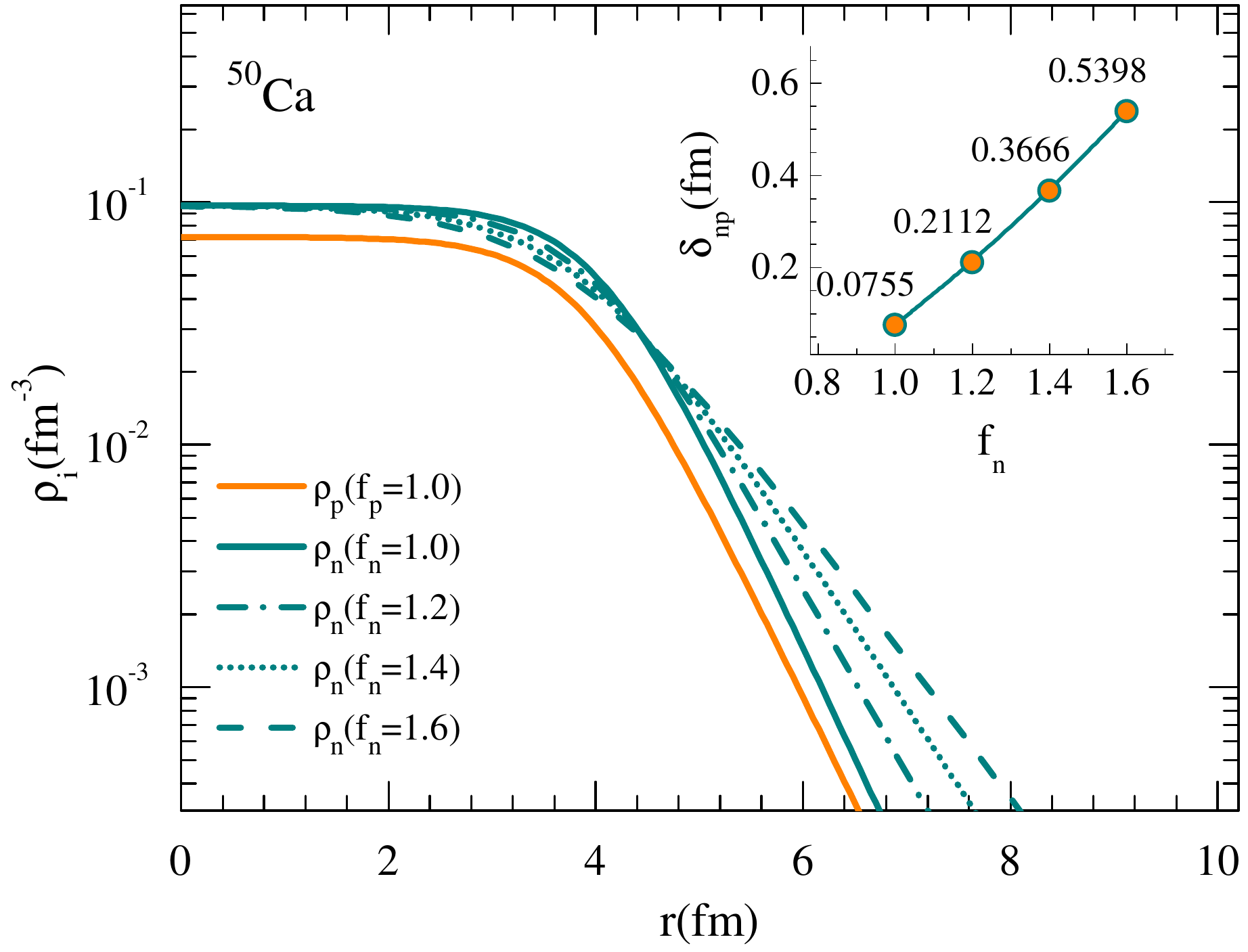}
    \caption{The proton and neutron density distributions of $^{50}$Ca computed from the droplet model. The correlation between $\delta_{np}$ and $f_n$ of $^{50}$Ca  is in the inset plot.}
    \label{f1density}
\end{figure}

In the phase space initialization of the projectile and target in the present IQMD model, the density distributions of protons and neutrons are distinguished from each other. The proton and neutron density distributions for the initial projectile and target nuclei are taken from the droplet model. By adjusting the diffuseness parameter of neutron density in the droplet model for the projectile, we can get different skin size in density distributions ~\cite{Sun2011nskin,Dai2014nskinPRC,Dai2015nskinPRC,Ma2011nskinCPC},
\begin{equation}\label{densitydroplet}
\rho_{i}(r) = \frac{\rho_{i}^{0}}{1+\exp(\frac{r-C_{i}}{f_{i}t_{i}/4.4})}, i = n, p, 
\end{equation}
where $\rho_{i}^{0}$ is the normalization constant which can ensure that the integration of the density distribution is equal to the number of protons ($i=p$) or neutrons ($i=n$), $C_i$ is half the density radius of the proton or neutron density distribution, and $f_i$ is introduced to adjust the diffuseness parameter $t_i$. More details can be found in Refs.~\cite{Sun2011nskin,Sun2011nskin2,Dai2014nskinPRC,Dai2015nskinPRC,Myers1983dropletNPA}. In this work, $f_p = 1.0$ is used in Eq. (\ref{densitydroplet}) for the proton density distribution, while we take $f_n = 1.0, 1.2, 1.4, 1.6$ in Eq.~(\ref{densitydroplet}) for the neutron density distributions in order to obtain different values of $\delta_{np}$. In Fig.~\ref{f1density}, we plot the proton and neutron density distributions of $^{50}$Ca computed from the droplet model. The related $\delta_{np}$ values of $^{50}$Ca are also included in the inset. It can be found that, with the increase of $f_n$, the neutron density distribution is more extended. Using these density distributions, the initial coordinates of nucleons in projectile and target nuclei are sampled via the Monte Carlo method. After IQMD initialization, the candidates of projectile and target nuclei are strictly selected by checking the stability of the sampled nuclei in the mean field. 

Two-body collision as one of three important components in intermediate-energy heavy-ion collisions, it is well known that in-medium effects cannot be ignored in this process, especially in the Fermi-energy range ~\cite{Wang2020photonEPJA,Wang2019corrPRC,Lopez2014nncsPRC,Wang2020nncsPLB}. To date, there are several available forms of the in-medium nucleon-nucleon cross section (in-medium NNCS) ~\cite{Li1993nncs,Danielewicz2002nncsAPPB,Coupland2011nncsPRC}. In the IQMD calculations, we take the screened cross section as the in-medium NNCS instead of the free nucleon-nucleon cross section (free NNCS) parameterized from  experimental measurements ~\cite{Chen1968nncsPR}. The formula is derived from the geometric reasoning that the geometric cross section radius cannot exceed the interparticle distance ~\cite{Danielewicz2002nncsAPPB,Coupland2011nncsPRC},
\begin{align}\label{inmediumnncs}
\sigma^{in-medium}_{NN} = &\sigma_{0}\tanh(\sigma^{free}_{NN}/\sigma_{0}),  \\
\sigma_{0} = &y\rho^{-2/3}, ~~~y = 0.85.
\end{align}
Here $\rho$ denotes the single-particle density. It can be seen that $\sigma^{in-medium}_{NN}$ is strongly dependent on the density of the scattered nucleons. In Ref.~\cite{Wang2020photonPRC}, the hard photon energy spectra from our calculations are compared with the experimental data, which indicates that the calculated results employing in-medium NNCS in the IQMD model are in good agreement with experimental results.

Considering that the procedure of Pauli blocking is another important component in intermediate-energy heavy-ion collision and the Pauli blocking effects in most QMD versions underestimate the blocking probability due to the fluctuations~\cite{zhang2018ComparisonPRC}, we performed some box calculations and confirmed that the Pauli blocking code in the present IQMD model is reasonable in our recent article~\cite{Wang2020photonPRC}.

\subsection{ Hard photon production probability}
\label{photonformula}
Hard photons in intermediate-energy heavy-ion collisions mainly originate from incoherent proton-neutron bremsstrahlung, i.e. $ p + n \to p + n + \gamma$. The elementary double-differential hard-photon production probability in the nucleon-nucleon center-of-mass frame employs the hard-sphere collision limit from Ref.~\cite{Jackson1962elebook} and is modified in Ref.~\cite{Cassing1986photonPLB} for energy conservation,
\begin{equation}\label{productionrate}
\frac{d^{2}P}{dE_{\gamma}d\Omega_{\gamma}} = \frac{\alpha_c}{12\pi^2}\frac{1}{E_{\gamma}}(2\beta^{2}_{f}+3sin^{2}\theta_{\gamma}\beta^{2}_{i}), 
\end{equation}
where $\alpha_c$ is the fine structure constant,  $ E_{\gamma} $ is the energy of the emitted photon, $ \beta_{i} $ and  $ \beta_{f} $ are the initial and final velocities of the proton, and $ \theta_{\gamma} $ is the angle between the momenta of the incident proton and the emitted photon.
 
\section{RESULTS AND DISCUSSION}
\label{resultsanddiscussion}
In the present work, the collisions of $^{50}$Ca and $^{40}$Ca projectiles with $^{40}$Ca and $^{12}$C targets, respectively, at incident energies ($E_{int}$) from $40$ to $150$ MeV/nucleon are simulated in the framework of the IQMD model with the in-medium NNCS in the process of two-body collisions. To investigate the neutron skin effects on hard photon emission in intermediate-energy heavy-ion collisions, we only focus on the central and peripheral collisions. For central collisions, the collision centrality takes $ 0\%-10\% $ and peripheral collisions correspond to $80\%-100\% $ centrality. Here, the centrality is defined by $\frac{100\pi b^{2}}{\pi b_{max}^{2}}$, where $b$ denotes impact parameter and $b_{max}$ is the summation of the radii of projectile and target nuclei. The direct hard photons which are emitted from incoherent proton-neutron bremsstrahlung at the earlier stage of the heavy-ion reaction should be more sensitive to the neutron skin thickness than thermal hard photons. That is the reason why we only check the effects of neutron skin thickness on the direct hard photon emission in this article.  It is important to note that the time evolution of the dynamical process in our calculation is simulated until $100$ fm/c, which is the separation time ($t_s$) between direct hard photons and thermal hard photons based on our recent work~\cite{Wang2020photonPRC,Wang2020photonEPJA}.

\subsection{Yield and yield ratio of direct hard photons}
\label{yield}

\begin{figure} 
    \includegraphics[width=8.6cm]{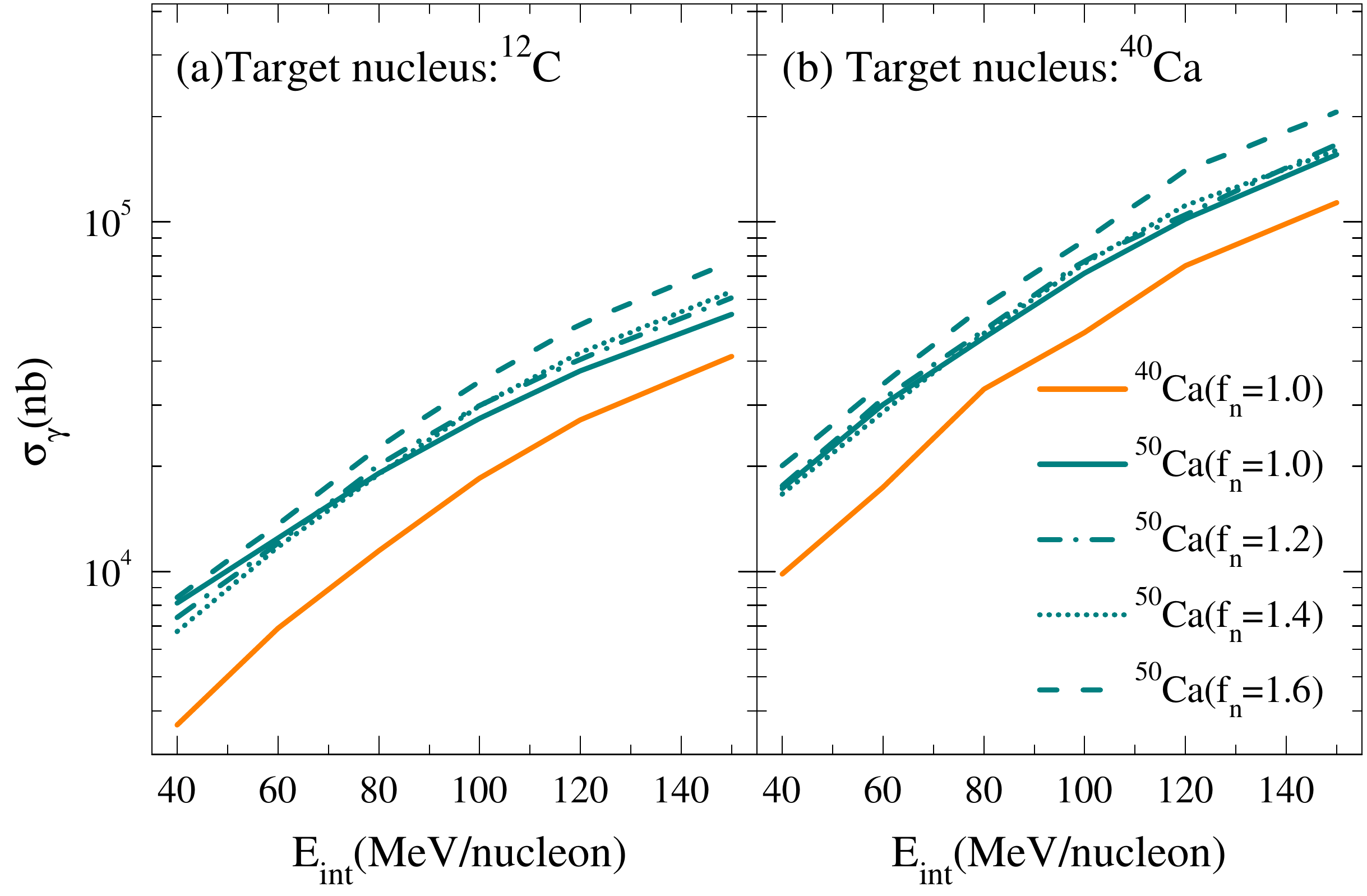}
    \caption{Total yield of direct hard photons from  peripheral collisions of $^{50}$Ca and $^{40}$Ca projectiles with $^{12}$C (a) and $^{40}$Ca (b) targets as a function of incident energy. Different lines correspond to the different neutron-skin thicknesses of the $^{50}$Ca projectile.}
    \label{f2yield}
\end{figure}

Figure ~\ref{f2yield} first plots the incident energy dependence of the direct hard photon yields considering the full rapidity range covered in peripheral collisions of $^{40}$Ca and $^{50}$Ca projectiles with $^{12}$C and $^{40}$Ca targets, respectively. It can be seen that there are more direct hard photons produced with the increase of $E_{int}$ from $40$ to $150$ MeV/nucleon. The result is consistent with that in Ref.~\cite{Wang2020photonPRC}. A comparison between reactions induced by $^{40}$Ca and $^{50}$Ca with $f_n=1.0$ shows that more direct hard photons are emitted from the reaction with increasing neutron excess of the projectile. Moreover, we perform a comparison with the total yield of direct hard photons produced from  the reactions induced by $^{50}$Ca with different $f_n$, which corresponds to different value of neutron skin thickness. With increasing neutron skin thickness, the total yield of direct hard photons will also increase for $E_{int}$ larger than $100$ MeV/nucleon. It indicates that a larger neutron skin thickness can enhance the opportunity of incoherent proton-neutron bremsstrahlung in peripheral collisions so that more direct hard photons are produced, which is in accordance with the results in Ref.~\cite{Wei2015photonPRC}. 

\begin{figure}
    \includegraphics[width=8.6cm]{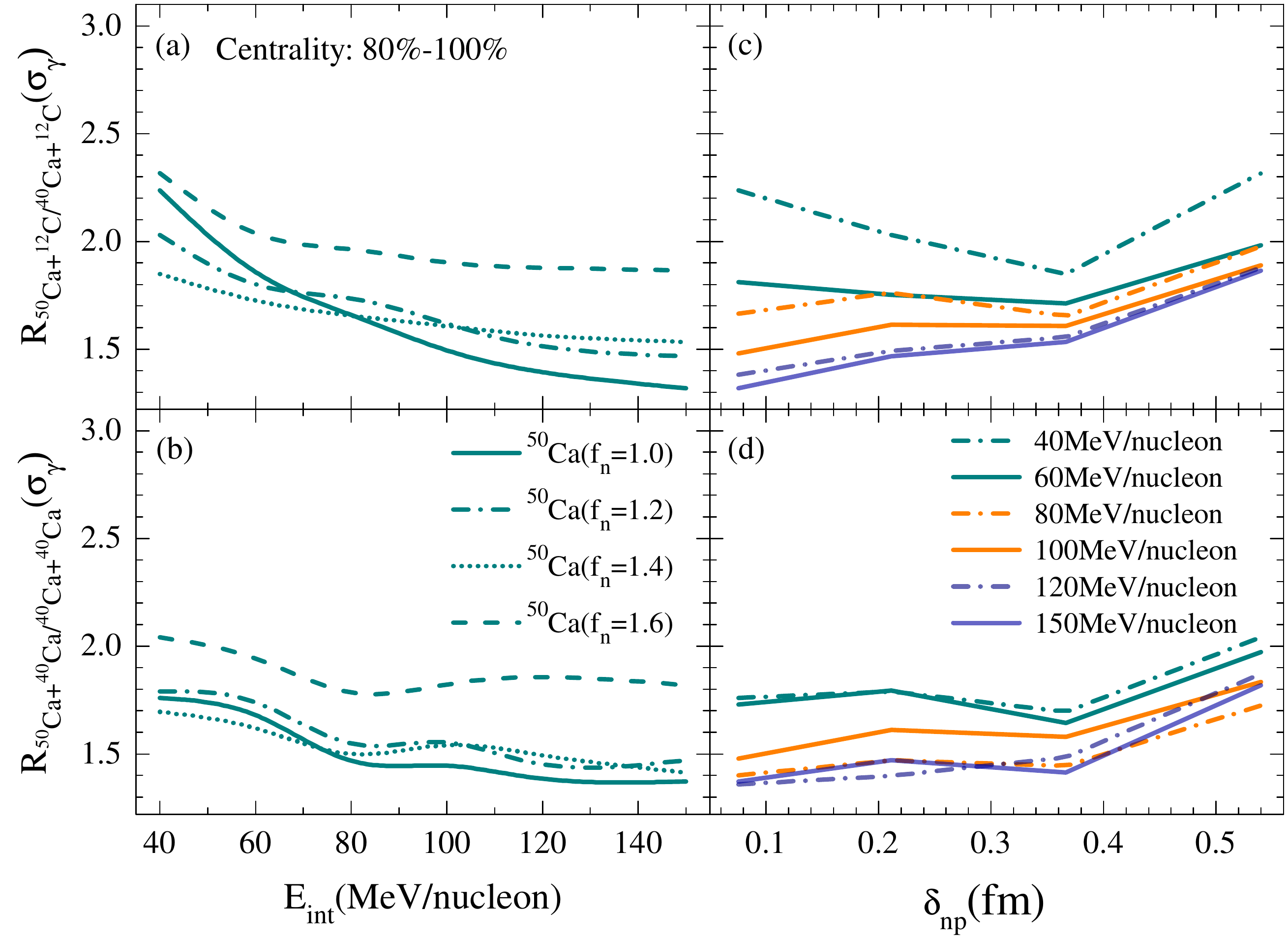}
    \caption{ Yield ratios of direct hard photons from two similar reactions for centrality $80\%-100\%$ as a function of incident energy [(a) and (b)] and neutron skin thickness [(c) and (d)].}
    \label{f3Ryield}
\end{figure}

In order to cancel out the systematic errors to some extent, we define the yield ratios of direct hard photons from two similar reactions to probe the neutron skin thickness by
\begin{align}
\label{R1sigma}
R_{^{50}{\rm Ca}+^{12}{\rm C}/^{40}{\rm Ca}+^{12}{\rm C}}(\sigma_{\gamma})=\frac{\sigma_{\gamma}(^{50}{\rm Ca}+^{12}{\rm C})}{\sigma_{\gamma}(^{40}{\rm Ca}+^{12}{\rm C})},   \\
R_{^{50}{\rm Ca}+^{40}{\rm Ca}/^{40}{\rm Ca}+^{40}{\rm Ca}}(\sigma_{\gamma})=\frac{\sigma_{\gamma}(^{50}{\rm Ca}+^{40}{\rm Ca})}{\sigma_{\gamma}(^{40}{\rm Ca}+^{40}{\rm Ca})},
\label{R2sigma}
\end{align}
which was also proposed in Ref.~\cite{Yong2008photonPLB}. Note that the reactions of $^{40}$Ca + $^{12}$C and $^{40}$Ca + $^{40}$Ca are used as referential reactions. Based on Eqs.~(\ref{R1sigma}) and ~(\ref{R2sigma}), the yield ratios of direct hard photons emitted from the two similar peripheral collisions as a function of incident energy and neutron skin thickness, respectively, are shown in Fig.\ref{f3Ryield}. Comparing with the calculated results from the reactions induced by the neutron-rich projectile $^{50}$Ca with different $f_n$ in Fig. \ref{f3Ryield} (a) and \ref{f3Ryield} (b), we see that the value of the yield ratio generally keeps rising with $f_n$ changing from $1.0$ to $1.6$ when $E_{int}$ is larger than $100$ MeV/nucleon, especially for $^{50}$Ca + $^{12}$C collisions. However, there appears an interesting phenomenon in the $^{50}$Ca+$^{40}$Ca collisions, i.e. the incident energy dependence of yield ratio between $f_n$ = $1.2$ and $f_n$ = $1.4$ has an inversion at about $E_{int}$ = 140 MeV/nucleon. The sensitivity of yield ratio to neutron skin size is reduced, probably by a mixing of different sources and mechanisms at higher incident energies, which can be disentangled with rapidity dependent analyses shown in Sec. III B. Furthermore, the neutron skin thickness dependence of the yield ratio from two similar reactions for centrality  $80\%-100\%$ can be observed clearly in Fig.~\ref{f3Ryield}(c) and \ref{f3Ryield}(d). The yield ratios of direct hard photons from two similar reactions show a monotonic increase tendency with the increasing the neutron skin thickness of $^{50}$Ca when $E_{int}$ is about $120$ MeV/nucleon. It indicates that the yield ratio of direct hard photons is sensitive to neutron skin thickness at an incident energy of about $120$ MeV/nucleon.

\begin{figure}
    \includegraphics[width=8.6cm]{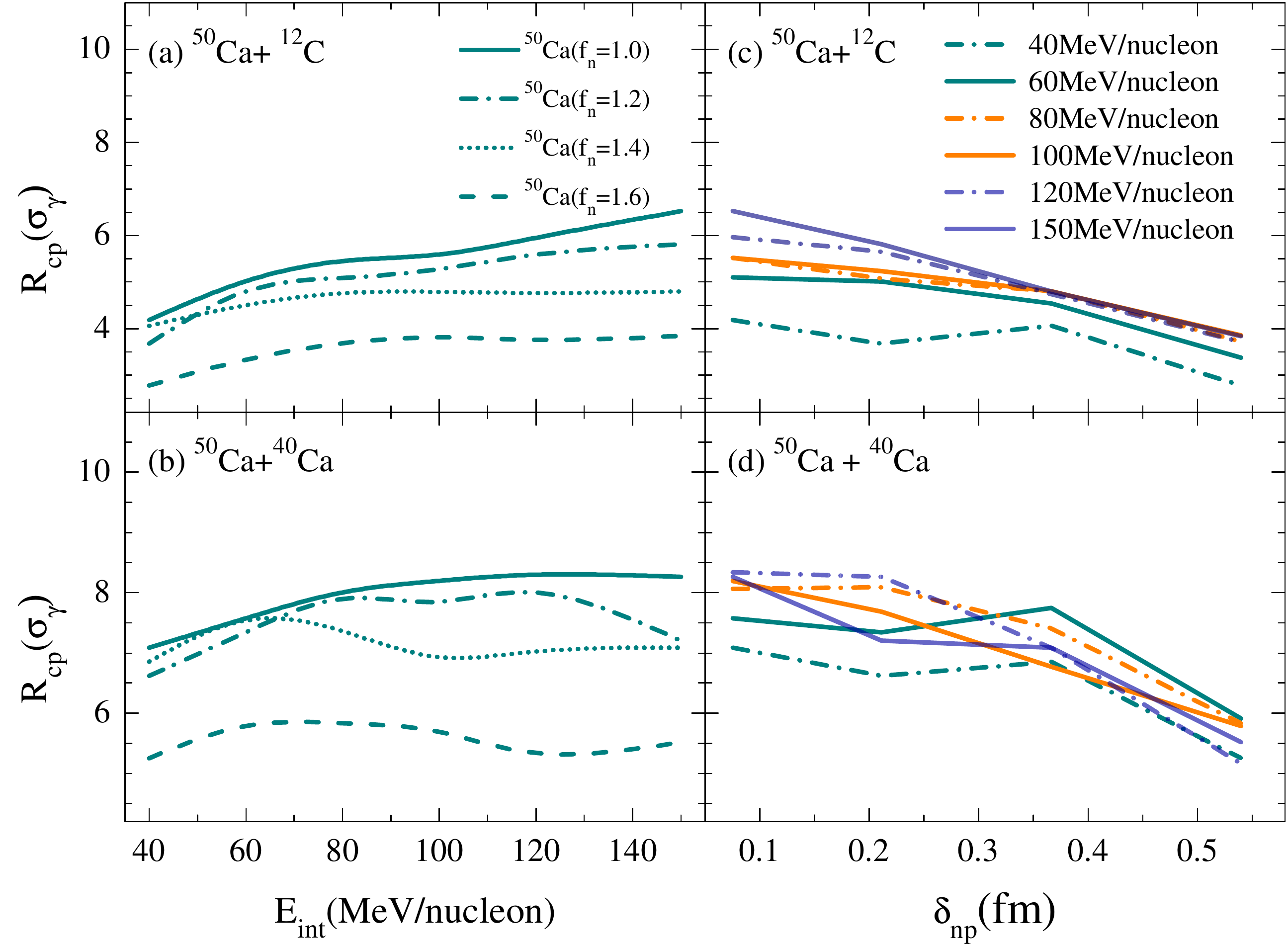}
    \caption{Yield ratios of direct hard photons from central and peripheral collisions in the same reaction systems of $^{50}$Ca + $^{12}$C [(a) and (c)] and $^{50}$Ca + $^{40}$Ca [(b) and (d)] as a function of incident energy [(a) and (b)] and neutron skin thickness [(c) and (d)]).}
    \label{f4Rcpyield}
\end{figure}

Similarly to the yield ratio of direct hard photons from the two reactions, we also check the effect of neutron skin thickness on direct hard photon production using the yield ratio of direct hard photons from the central and peripheral collisions in the same reaction system by the following formula
\begin{equation}\label{Rcpsigma}
R_{cp}(\sigma_{\gamma}) = \frac{\sigma_{\gamma}(Central~coll.)}{\sigma_{\gamma}(Peripheral~coll.)}.
\end{equation}
In the above equation, the numerator is evaluated for $0\%-10\%$ centrality and the denominator is computed for $80\%-100\%$ centrality in our calculations. The incident energy and neutron skin thickness dependence of $R_{cp}$ in $^{50}$Ca + $^{12}$C and $^{50}$Ca + $^{40}$Ca collisions is shown in Fig.~\ref{f4Rcpyield}. It is clear that the value of $R_{cp}$ has a tendency to decrease with increasing $f_n$ except at lower incident energy because more direct hard photons are emitted from the peripheral collision induced by $^{50}$Ca with a larger neutron skin size. Meanwhile, we find that the $R_{cp}$ of direct hard photons originating from  $^{50}$Ca + $^{12}$C collisions is more sensitive to neutron skin thickness.

\subsection{Rapidity dependence of the direct hard photon multiplicity}
\label{multiplicity}

\begin{figure}
    \includegraphics[width=8.6cm]{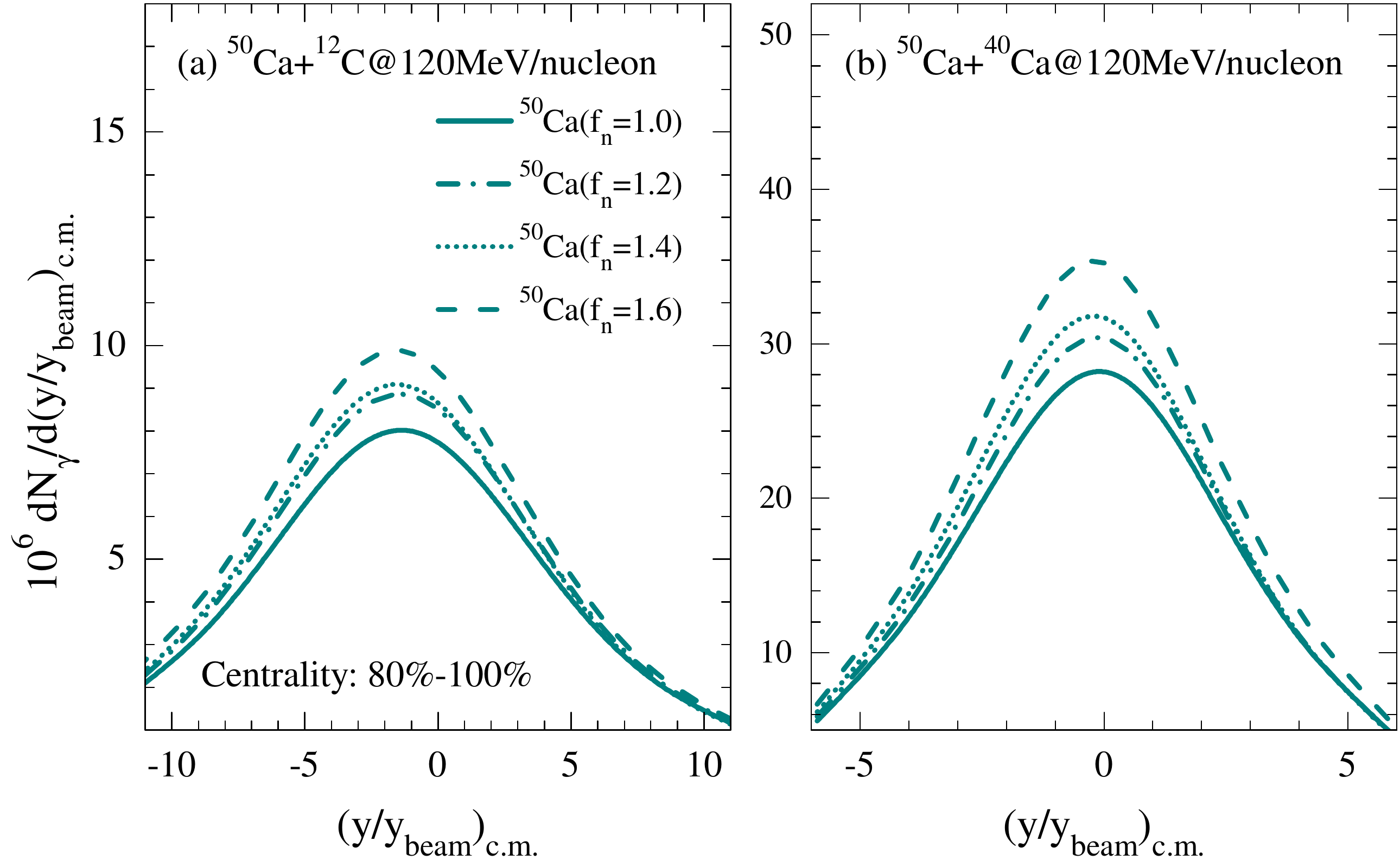}
    \caption{ Multiplicity of direct hard photon as a function of the reduced c.m. rapidity in the reactions of $^{50}$Ca + $^{12}$C (a) and $^{50}$Ca + $^{40}$Ca (b) for centrality $80\%-100\%$ and at a beam energy of 120 MeV/nucleon. Different lines correspond to $^{50}$Ca projectiles with different $f_n$.}
    \label{f5dNybeam}
\end{figure}

Based on the discussion in Sec.~\ref{yield}, the energy of about $120$ MeV/nucleon for incident nucleus $^{50}$Ca is a good reaction condition to probe the neutron skin effect on direct hard photon production.  So Fig.~\ref{f5dNybeam} only shows the rapidity distributions of the multiplicity of direct hard photons in collisions of $^{50}$Ca projectiles with $^{12}$C and $^{40}$Ca at an incident energy of $120$ MeV/nucleon and for centrality $80\%-100\%$. We find that the rapidity distributions of the multiplicity for direct hard photons are appreciably  sensitive to the neutron skin thickness. Meanwhile, we find that the multiplicities of direct hard photon show an increasing trend with the increase of neutron skin thickness of the neutron-rich projectile  $^{50}$Ca, especially at midrapidity from the participant region. The results also confirm that there are more direct hard photons emitted from the peripheral collisions induced by a projectile with a larger neutron skin size.

\begin{figure}
    \includegraphics[width=8.6cm]{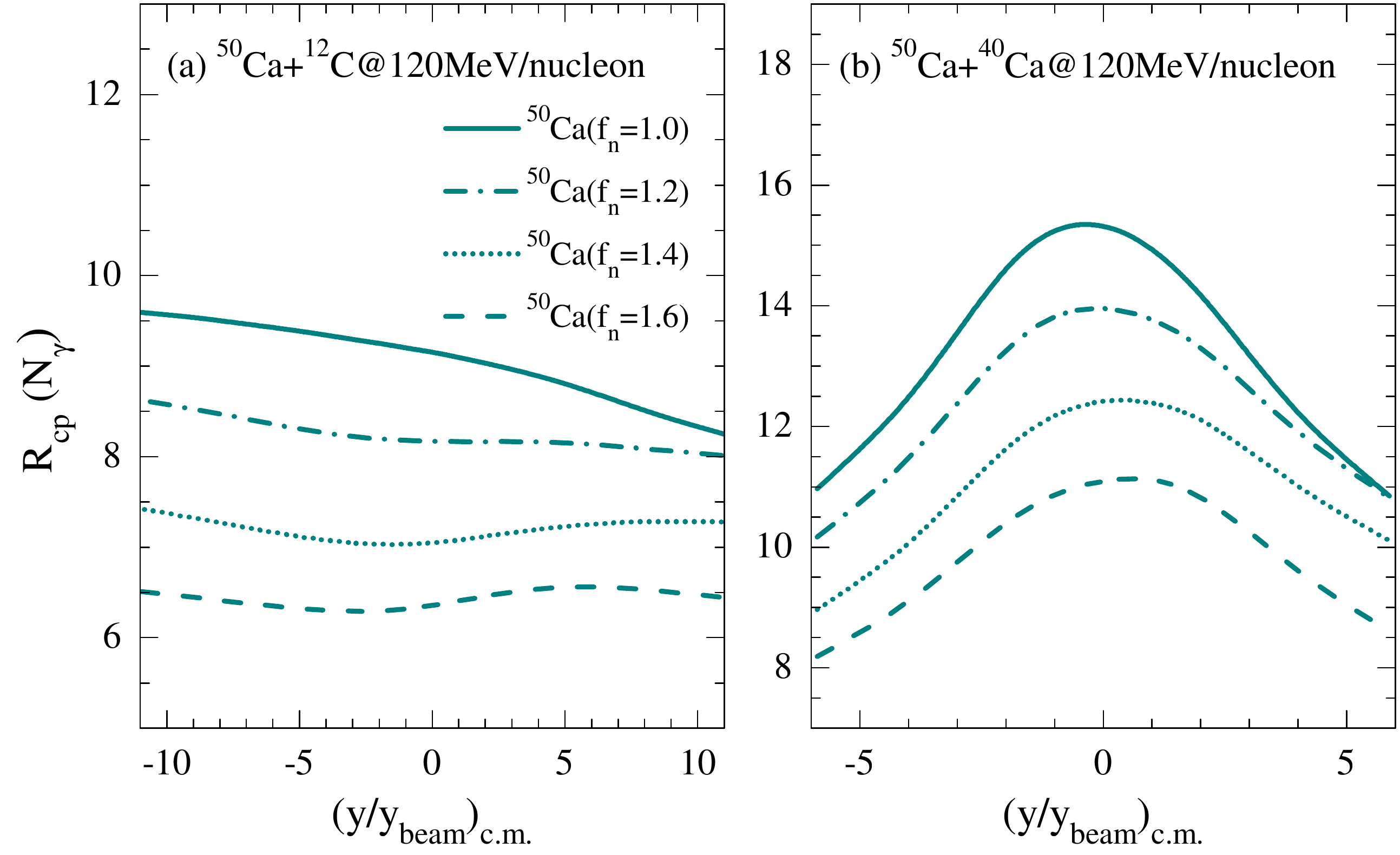}
    \caption{ The ratio of direct hard photon multiplicity from central and peripheral collisions in the same reaction systems of $^{50}$Ca+$^{12}$C (a) and $^{50}$Ca+$^{40}$Ca (b) as a function of the reduced c.m. rapidity. Different lines correspond to $^{50}$Ca projectiles with different $f_n$.}
    \label{f6RcpdNybeam}
\end{figure}

To cancel out the errors inside the reaction systems, we also employ the multiplicity ratio of direct hard photons emitted from  central and peripheral collisions in the  same reaction system, which reads 
\begin{equation}\label{RcpMult}
R_{cp}(N_{\gamma}) = \frac{dN_{\gamma}/d(y/y_{beam})_{c.m.}(Central~coll.)}{dN_{\gamma}/d(y/y_{beam})_{c.m.}(Peripheral~coll.)},
\end{equation}
where the collision centrality in the numerator and denominator is the same as that in Eq.~(\ref{Rcpsigma}). Figure \ref{f6RcpdNybeam} shows the calculated results from the reactions of $^{50}$Ca + $^{12}$C and $^{50}$Ca + $^{40}$Ca.  We see that the values of multiplicity ratio tend to decrease with increasing $f_n$, indicating that the ratio is also greatly sensitive to the neutron skin thickness. Here it is noted that $R_{cp}(N_{\gamma})$ for two systems behave differently as a function of reduced rapidity, which is because the reaction system $^{50}$Ca + $^{40}$Ca is much more symmetric than $^{50}$Ca + $^{12}$C, and we then see a peak of $R_{cp}(N_{\gamma})$ around $y/y_{beam}$ = 0 for $^{50}$Ca + $^{40}$Ca. However, the peak of  the rapidity dependence of $R_{cp}(N_{\gamma})$ in $^{50}$Ca + $^{12}$C collisions appears at a larger negative $y/y_{beam}$ near the target nucleus side.

In the present work, we clearly find that the rapidity distributions of both multiplicity and multiplicity ratio of direct hard photons emitted from the peripheral collisions, i.e., $dN_{\gamma}/d(y/y_{beam})c.m.$ and $R_{cp}(N_{\gamma})$, are sensitive to the neutron skin thickness of projectile nucleus $^{50}$Ca. Considering that the density slope parameter $L(\rho)$ of the nuclear symmetry energy $E_{sym}(\rho)$ exhibits a particularly strong positive correlation with the neutron skin thickness of heavy nuclei \cite{Chen2005PRC}, it indicates that direct hard photons may be also a good probe of $L(\rho)$. For the sensitivities of photon production to the compressibility and symmetry potentials, Ma {\it et al.}  have investigated them for inclusive events covering  all collision centralities in the framework of the Blotzmann-Uehling-Ulenbeck model and found that direct hard photon production weakly depends on the EOS and symmetry energy since direct hard photons are produced in the early stage of collisions so that they do not have enough time to feel the EOS influence \cite{Ma2012photonPRC}. Actually, photon production  dynamics is dominated by nucleon-nucleon collisions rather than the nuclear mean field, and effects of the symmetry energy on photons are then expected to become smaller. However, using the ratio of hard photon spectra  from two reactions of  isotopes, they found that the ratio seems sensitive to the symmetry energy in a previous study \cite{Yong2008photonPLB}. In the present work, we only focus on the peripheral collisions to explore the neutron skin thickness effects on direct hard photon emission. The dependence of direct hard photons produced from the peripheral collisions on the EOS and symmetry energy could be stronger than that from inclusive events, which need to be further studied in the near future.

From the experimental viewpoint, hard photons induced by neutron-rich projectiles can be measured by radioactive ion beam facilities such as RIBF at RIKEN, FRIB at MSU as well as HIAF at Huizhou \cite{HIAF,HIAF2} in the near future, which can access neutron-rich nuclei in the middle and heavy mass region with enough beam intensity. 
This kind of experiment can be integrated with experiments that aim to study the properties of radioactive nuclei and share beam time because the detectors for hard photons do not need to work in vacuum, and can be arranged outside the reaction target chamber. The forward emission of hard photons in the laboratory system can reduce the requirements for detector solid angle coverage.

\section{Summary}
\label{summary}
In summary, we have carried out a systematic study of the effects of neutron skin thickness on direct hard photon emissions from the reactions of $^{50}$Ca + $^{12}$C and $^{50}$Ca + $^{40}$Ca in the framework of the IQMD model. By adjusting the diffuseness parameter of neutron density in the droplet model for the projectile $^{50}$Ca to obtain different neutron skin thicknesses, the sensitivities of several observable to neutron skin size are explored. We find that the yield ratio of direct hard photons between central and peripheral collisions in the same reaction system, i.e., $R_{cp}(\sigma_{\gamma})$,  are more sensitive to neutron skin thickness than the yield $\sigma_{\gamma}$ and yield ratio between two similar reactions. We also study on the rapidity distribution of multiplicity $N_{\gamma}$ and multiplicity ratio $R_{cp}(N_{\gamma})$ of direct hard photons, and discover that both probes display appreciable sensitivity to neutron skin thickness. Meanwhile, we find that there are more direct hard photons produced  with the increase of neutron skin thickness in peripheral collisions. These results indicate that direct hard photons can be treated as an experimental observable to extract information on neutron skin thickness. 

Finally, we point out that  we have not considered the effects of the $\alpha$ clustering structure of $^{12}$C on hard photon emission in this work,, which could  play an additional role as shown in the framework of an extended quantum molecular dynamics model by Shi and Ma \cite{Shi2}, where they found that collective flows of direct photons are sensitive to the initial $\alpha$ clustering configuration. A future work in this direction deserves consideration. 

\vspace{.5cm}
This work is supported by the National Natural Science Foundation of China under Contracts No. 11890710, No.11890714, No.12147101, No. 12047514, No. 11875066, No. 11925502, No. 11961141003, No. 11935001 and No. 12105053，by the Strategic Priority Research Program of CAS under Grant No. XDB34000000, by the National Key R\&D Program of China under Grants No. 2016YFE0100900 and No. 2018YFE0104600, and by the Guangdong Major Project of Basic and Applied Basic Research No. 2020B0301030008.


\end{CJK}
\end{document}